\documentclass[prl,twocolumn,superscriptaddress]{revtex4-1}
\usepackage{amsmath}
\usepackage{amssymb}
\usepackage{graphics}
\usepackage{graphicx}
\usepackage{color}
\usepackage{hyperref}
\newcounter{fnnumber}

\begin{document}
\title{Unconventional Surface Critical Behaviors Induced by Quantum Phase Transition from Two-Dimensional Affleck-Kennedy-Lieb-Tasaki Phase to N\'eel Order}

\author{Long Zhang}
\affiliation{International Center for Quantum Materials, School of Physics, Peking University, Beijing, 100871, China}

\author{Fa Wang}
\affiliation{International Center for Quantum Materials, School of Physics, Peking University, Beijing, 100871, China}
\affiliation{Collaborative Innovation Center of Quantum Matter, Beijing, 100871, China}

\begin{abstract}
A symmetry-protected topological phase has nontrivial surface states in the presence of certain symmetries, which can either be gapless or be degenerate. In this work, we study the physical consequence of such gapless surface states at the bulk quantum phase transition (QPT) that spontaneously breaks these symmetries. The two-dimensional Affleck-Kennedy-Lieb-Tasaki phase on a square lattice and its QPTs to N\'eel ordered phases are realized with the spin-$1/2$ Heisenberg model on a decorated square lattice. With large-scale quantum Monte Carlo simulations, we show that even though the bulk QPTs are governed by the conventional Landau phase transition theory, the gapless surface state induces unconventional universality classes of the surface critical behaviors.
\end{abstract}
\date{\today}
\maketitle


\emph{Introduction.---}Symmetry-protected topological (SPT) phases are bulk-gapped phases that cannot continuously evolve into direct product states without explicitly breaking certain symmetries or closing the bulk energy gap \cite{Gu2009, Chen2012a}. It is the generalization of the Haldane phase of spin-1 antiferromagnetic (AF) Heisenberg chains \cite{Haldane1983a, Haldane1983, Haldane1985} and the topological band insulators \cite{Hasan2010, Qi2011}. One of the hallmark of SPT phases is the presence of nontrivial surface states that are either gapless or degenerate, which cannot be removed if the symmetries are preserved. In this work, we explore the physical consequence of the gapless surface state when the bulk SPT phase undergoes a quantum phase transition (QPT) that spontaneously breaks the protecting symmetries.

\begin{figure}[b]
\centering
\includegraphics[width=0.48\textwidth]{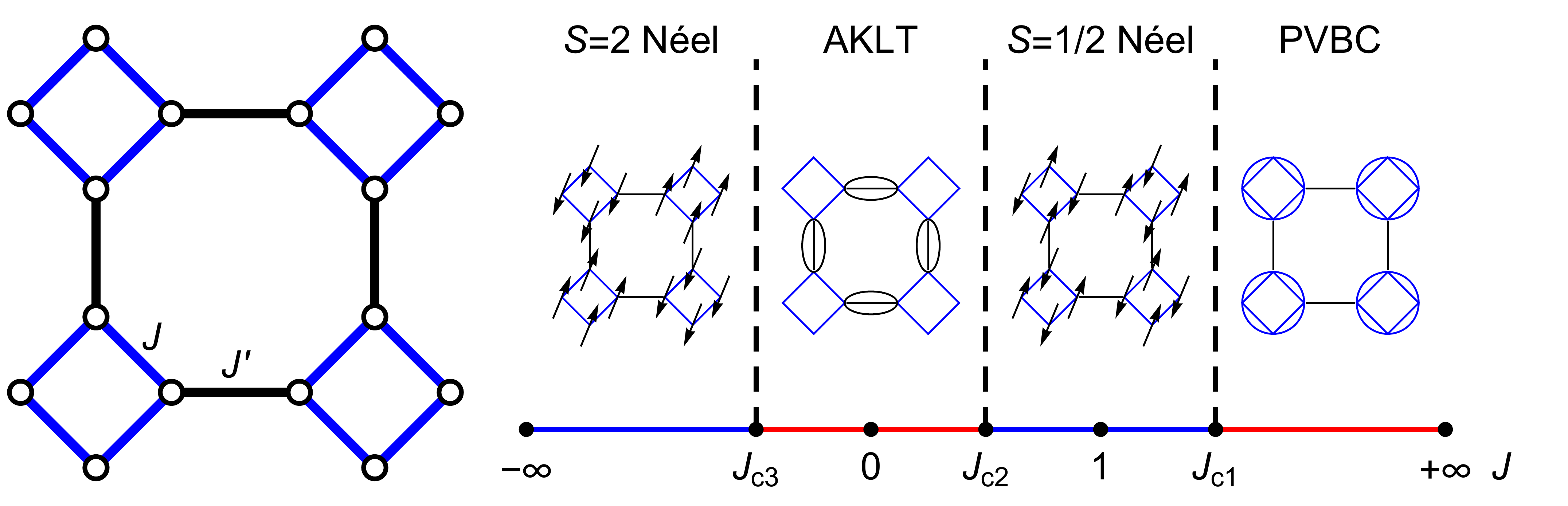}
\caption{Left: The decorated square lattice. Within each UC there are four sites. The inter-UC Heisenberg coupling $J'$ is antiferromagnetic and is set to be unity. The intra-UC Heisenberg coupling $J$ can be either ferromagnetic or antiferromagnetic. Right: The quantum phase diagram obtained by tuning $J$. Four phases are separated by three QCPs. A cartoon is sketched for a representative state in each phase. The ellipses in the AKLT phase and the circles in the PVBC phase denote the spin singlets formed on the bonds and on the plaquettes, respectively. Note that the ordered patterns are different in the two N\'eel ordered phases.}
\label{fig:lattice}
\end{figure}

We study the spin-$1/2$ Heisenberg model on a decorated square lattice (also called a square-octagon lattice, see Fig. \ref{fig:lattice}, left panel). It realizes the two-dimensional (2D) spin-$2$ Affleck-Kennedy-Lieb-Tasaki (AKLT) phase \cite{Affleck1987a}, which is an SPT phase protected by the spin rotational symmetry together with the spatial translational symmetry \cite{Chen2011b}. Its surface state is a spin-$1/2$ chain with effective AF Heisenberg interactions, which is gapless if the symmetries are preserved. Besides the AKLT phase, this model can also be tuned by varying the intra-unit cell (UC) coupling strength $J$ into two different N\'eel ordered phases, which spontaneously break the spin rotational symmetry, and a topologically trivial disordered phase dubbed a plaquette valence bond crystal (PVBC) (Fig. \ref{fig:lattice}, right panel). These phases are separated by three quantum critical points (QCP). Two of them are from the AKLT phase to the N\'eel ordered phases, and the other one is from the PVBC phase to one of the N\'eel ordered phases. Therefore, this model is suitable for examining the role of the symmetry-protected gapless surface state at the QCP and contrasting it with the more conventional case.

This model is free of magnetic frustration and can be studied with quantum Monte Carlo (QMC) simulations. Two QCPs on the AF side ($J>0$) were identified before \cite{Troyer1996}, and one of them, $J_{c1}$ from the PVBC phase to the $S=1/2$ N\'eel order, was shown to be in the three-dimensional (3D) O(3) universality class \cite{Troyer1997}, which is expected from the Landau phase transition theory of spontaneous symmetry breaking. In this work, we carry out large-scale QMC simulations around all three QCPs and show that all of them belong to the 3D O(3) universality class no matter whether they are adjacent to the AKLT SPT phase or not. Therefore, the SPT order does not change the universality class of the bulk QCP that spontaneously breaks its protecting symmetry.

What is the physical consequence of the gapless surface state of an SPT phase at the QCP then? If a disordered system with free surfaces undergoes a phase transition to a symmetry-breaking ordered phase, the long-range order is also induced on the surface. At the bulk critical point, the surface also has long-range correlation and exhibits singularities in its physical quantities, which are called surface critical behaviors. If the surface is gapped in the disordered phase, which is the case of an ``ordinary transition'', its long-range correlation at the critical point is purely induced from the bulk; therefore, the surface critical behaviors are expected to be fully determined by the bulk. However, if an SPT phase undergoes a phase transition that breaks its protecting symmetry, the gapless surface state hybridizes with the gapless modes at the critical point and leads to unconventional surface critical behaviors.

Surface critical behaviors have been studied in classical phase transitions since more than four decades ago \cite{Binder1983}. They are characterized by several critical exponents, which are defined with the surface thermodynamic quantities and correlation functions and satisfy scaling relations among each other and with the bulk critical exponents \cite{Barber1973, Binder1983}. We show that the surface critical exponents at the PVBC-N\'eel order QCP are consistent with those found in the 3D classical Heisenberg model \cite{Deng2005}, which confirms the universality. At the other two QCPs from the AKLT phase to the N\'eel ordered phases, the surface critical exponents are found to be distinct from those at the PVBC-N\'eel order QCP despite the same bulk universality class. These exponents satisfy the scaling relations, thereby consistently establishing a new surface universality class at each QCP, which is the physical consequence of the gapless surface state.

\emph{Model and methods.---}The spin-$1/2$ Heisenberg model on the decorated square lattice (Fig. \ref{fig:lattice}, left panel) is given by
\begin{equation}
\label{eq:h}
H=J\sum_{\langle ij\rangle}\mathbf{S}_{i}\cdot \mathbf{S}_{j}+J'\sum_{\langle ij\rangle'}\mathbf{S}_{i}\cdot \mathbf{S}_{j},
\end{equation}
in which $\langle ij\rangle$ denotes the intra-UC bonds, while $\langle ij\rangle'$ denotes the inter-UC bonds. We consider the AF inter-UC coupling and set $J'$ to be unity. By tuning the intra-UC coupling $J$ from antiferromagnetic to ferromagnetic, the model realizes four phases separated by three QCPs denoted by $J_{ci}$, $i=1,2,3$ (Fig. \ref{fig:lattice}, right panel). The nature of each phase can be understood by examining one representative point:

(a) $J\rightarrow +\infty$: In this limit, the lattice reduces to disjoint plaquettes. The ground state is the direct product state of the spin singlets formed on these plaquettes, which is disordered and gapped. Therefore, we dub this phase a ``plaquette valence bond crystal''.

(b) $J = 1$: This model is equivalent to the nearest-neighbor AF Heisenberg model on the CaV$_{4}$O$_{9}$ lattice. Its ground state has N\'eel order \cite{Troyer1996}.

(c) $J = 0$: The lattice reduces to disjoint inter-UC bonds. Its ground state is the direct product state of these bond singlets, which is disordered and gapped. Projecting this state into the total-spin-$2$ subspace in each UC precisely yields the 2D AKLT state on the square lattice \cite{Affleck1987a}. On a lattice with a free straight surface, the dangling bonds on the surface form a spin-$1/2$ chain with an effective AF coupling, which is gapless if the spin rotational symmetry and the translational symmetry are preserved. Therefore, we believe that this state is adiabatically connected to the 2D AKLT state.

(d) $J\rightarrow -\infty$: Because of the dominant intra-UC ferromagnetic coupling, each UC has total spin $2$. These blocks form N\'eel order due to the inter-UC AF coupling.

We study all three QCPs in detail using the stochastic series expansion (SSE) QMC with the loop algorithm \cite{sandvik1991, sandvik1999} on lattices with the linear size $L$ ($4L^{2}$ sites) from $8$ to $80$ and the inverse temperature $\beta=2L$. The fully periodic boundary condition is adopted to extract the QCP positions and the bulk critical exponents; afterwards the open boundary condition is taken in one direction to study the surface critical behaviors. For each lattice size, $10^{6}$ Monte Carlo steps are performed at each coupling strength. Various physical quantities are measured after each step. The parallel tempering and the multihistogram reweighting techniques are adopted to further improve the data quality. The statistical errors are found to be negligible in fitting the critical exponents.

\emph{Bulk QCPs.---}All three QCPs are associated with the spontaneous breaking of the spin rotational symmetry. The order parameter is the staggered magnetization,
\begin{equation}
m_{s}^{z}=\frac{1}{4L^{2}}\sum_{i}(-1)^{i}S_{i}^{z},
\end{equation}
in which $(-1)^{i}=+1$ or $-1$ according to the N\'eel order patterns sketched in Fig. \ref{fig:lattice}. The Binder ratios are derived from $m_{s}^{z}$, $Q_{1}=\langle (m_{s}^{z})^{2}\rangle / \langle |m_{s}^{z}|\rangle^{2}$, and $Q_{2}=\langle (m_{s}^{z})^{4}\rangle/\langle (m_{s}^{z})^{2}\rangle^{2}$. The second-moment correlation length is derived from the static spin structure factors, $S(\mathbf{q})=\sum_{\mathbf{r}}e^{-i\mathbf{q}\cdot\mathbf{r}}C(\mathbf{r})$,
\begin{equation}
\label{eq:xi}
\xi=\frac{L}{2\pi}\sqrt{\frac{S(\pi,\pi)}{S(\pi+2\pi/L,\pi)}-1},
\end{equation}
in which $C(\mathbf{r})=\langle S_{\mathbf{r}}^{z}S_{0}^{z}\rangle$ is the spin correlation function. The spin stiffness $\rho_{s}$ and the uniform susceptibility $\chi_{u}$ are measured with the improved estimators \cite{Sandvik2010}. At a QCP of the AF Heisenberg model, $Q_{1}$, $Q_{2}$, $\xi/L$, $\rho_{s}\beta$ and $\chi_{u}\beta$ are dimensionless, so these quantities are adopted to estimate the QCPs with the standard $(L,2L)$ crossing analysis \footnote{See Supplemental Materials for details of the finite-size scaling analysis of the bulk QCPs.}\setcounter{fnnumber}{\thefootnote}. The results are listed in Table \ref{tab:exponents}.

\begin{table*}[t]
\centering
\caption{Summary of the bulk and the surface critical exponents. The bulk critical exponents of the 3D O(3) field theory calculated by high-order perturbation \cite{Guida1998} and the surface critical exponent $y_{h_{1}}$ of the 3D classical Heisenberg model estimated by Monte Carlo simulations \cite{Deng2005} are also listed for comparison.}
\label{tab:exponents}
\begin{tabular}{ccccccccc}
\hline \hline
         	& $J_{c}$         	& $z$          	& $\nu$        	& $\eta$       	& $\beta$ 		& $y_{h_{1}}$ 	& $\eta_{\parallel}$ & $\eta_{\perp}$ \\
\hline
$J_{c1}$ 	& $1.064382(13)$  	& $1.0008(16)$ 	& $0.7060(13)$ 	& $0.0357(13)$ 	& $0.3663(8)$  	& $0.810(20)$  	& $1.327(25)$ 		 & $0.680(8)$     \\
$J_{c2}$ 	& $0.603520(10)$  	& $1.001(5)$   	& $0.7052(9)$  	& $0.031(4)$   	& $0.3642(13)$ 	& $1.7276(14)$ 	& $-0.449(5)$ 		 & $-0.2090(15)$  \\
$J_{c3}$ 	& $-0.934251(11)$ 	& $0.9999(13)$ 	& $0.7052(15)$ 	& $0.0365(10)$ 	& $0.3659(9)$  	& $1.7802(16)$ 	& $-0.561(4)$ 		 & $-0.2707(24)$  \\
3D O(3) field theory \cite{Guida1998} & --		& --        	& $0.7073(35)$ 	& $0.0355(25)$ 	& $0.3662(25)$ 	& 				&					 &\\
3D classical Heisenberg \cite{Deng2005}  & -- 	& --			&				&				&				& $0.813(2)$	&					 &\\
\hline \hline
\end{tabular}
\end{table*}

Various physical quantities, including the slopes of the above dimensionless quantities, the staggered magnetic susceptibility $\chi_{s}$, the static spin structure factor $S(\pi, \pi)$, the spin correlation at half of the lattice size $C(L/2,L/2)$, and the staggered magnetization $m_{s}^{z}$, are evaluated at the QCPs with the reweighting technique. They are used in the finite-size scaling analysis to obtain the critical exponents \footnotemark[\thefnnumber]. The dynamical critical exponent $z$ is found to be $1$ within error bars at all three QCPs \footnotemark[\thefnnumber], which is consistent with the asymptotic Lorentz invariance. The results of the correlation length exponent $\nu$, the anomalous dimension $\eta$, and the magnetization exponent $\beta$ are summarized in Table \ref{tab:exponents}. Results of the 3D O(3) field theory \cite{Guida1998} are also listed for comparison. All three QCPs are found to be consistent with the 3D O(3) universality class, which is expected from the Landau phase transition theory.

\emph{Surface critical behaviors.---}For a system with two free surfaces, the singular part of the free energy density in the quantum critical regime is contributed by the bulk part and the surface part \cite{Binder1983, Barber1983},
\begin{equation}
f(\delta, h, h_{1}; L) = f_{b}(\delta, h; L) + \frac{2}{L} f_{1}(\delta, h, h_{1}; L),
\end{equation}
in which $\delta = J-J_{c}$. $h$ is the staggered magnetic field that couples to the bulk order parameter $m_{s}^{z}$, while $h_{1}$ is the surface staggered field that couples to the order parameter restricted to the surface.

The bulk free energy $f_{b}$ satisfies the scaling ansatz on a finite lattice with a linear size $L$,
\begin{equation}
f_{b}(\delta, h; L) \sim L^{-(d+z)} \tilde{f}_{b}(\delta L^{y_{\delta}}, h L^{y_{h}}),
\end{equation}
in which $\tilde{f}_{b}$ is a smooth function of its arguments. $y_{\delta}$ and $y_{h}$ are the scaling dimensions of $\delta$ and $h$, respectively. They are related to $\nu$ and $\beta$ by $\nu=1/y_{\delta}$, and $\beta/\nu=d+z-y_{h}$.

The surface free energy $f_{1}$ satisfies a similar scaling ansatz,
\begin{equation}
\label{eq:fs}
f_{1}(\delta, h, h_{1}; L) \sim L^{-(d+z-1)} \tilde{f}_{1}(\delta L^{y_{\delta}}, h L^{y_{h}}, h_{1} L^{y_{h1}}),
\end{equation}
in which the scaling dimension $y_{h_{1}}$ of the surface field $h_{1}$ enters as an independent exponent to characterize the universality class of the surface critical behaviors.

The surface critical behaviors can be derived from Eq. (\ref{eq:fs}). The surface staggered magnetic susceptibility $\chi_{1,1}$ with respect to the surface field $h_{1}$ has the following finite-size scaling form,
\begin{equation}
\label{eq:chi11}
\chi_{1,1}=-\frac{\partial^{2}f_{1}}{\partial h_{1}^{2}}\sim L^{-(d+z-1-2y_{h_{1}})}.
\end{equation}

The long-range spin correlation on the surface is characterized by two anomalous dimensions, $\eta_{\parallel}$ and $\eta_{\perp}$, which are defined as follows. With one endpoint denoted by $0$ fixed on the surface, the spin correlation function $C_{\parallel}(\mathbf{r})=\langle S_{\mathbf{r}}^{z}S_{0}^{z} \rangle$ with $\mathbf{r}$ parallel to the surface scales as
\begin{equation}
\label{eq:cparallel}
|C_{\parallel}(\mathbf{r})|\sim r^{-(d+z-2+\eta_{\parallel})},
\end{equation}
while $C_{\perp}(\mathbf{r})=\langle S_{\mathbf{r}}^{z}S_{0}^{z} \rangle$ with $\mathbf{r}$ perpendicular to the surface scales as
\begin{equation}
\label{eq:cperp}
|C_{\perp}(\mathbf{r})|\sim r^{-(d+z-2+\eta_{\perp})}.
\end{equation}

These surface critical exponents are not independent of each other. They satisfy the following scaling relations \cite{Barber1973, Lubensky1975, Binder1983},
\begin{align}
1-\eta_{\parallel} &=-(d+z-1-2y_{h_{1}}), \label{eq:relation1}\\
2\eta_{\perp} &=\eta_{\parallel}+\eta. \label{eq:relation2}
\end{align}

\begin{figure}[b]
\centering
\includegraphics[width=0.48\textwidth]{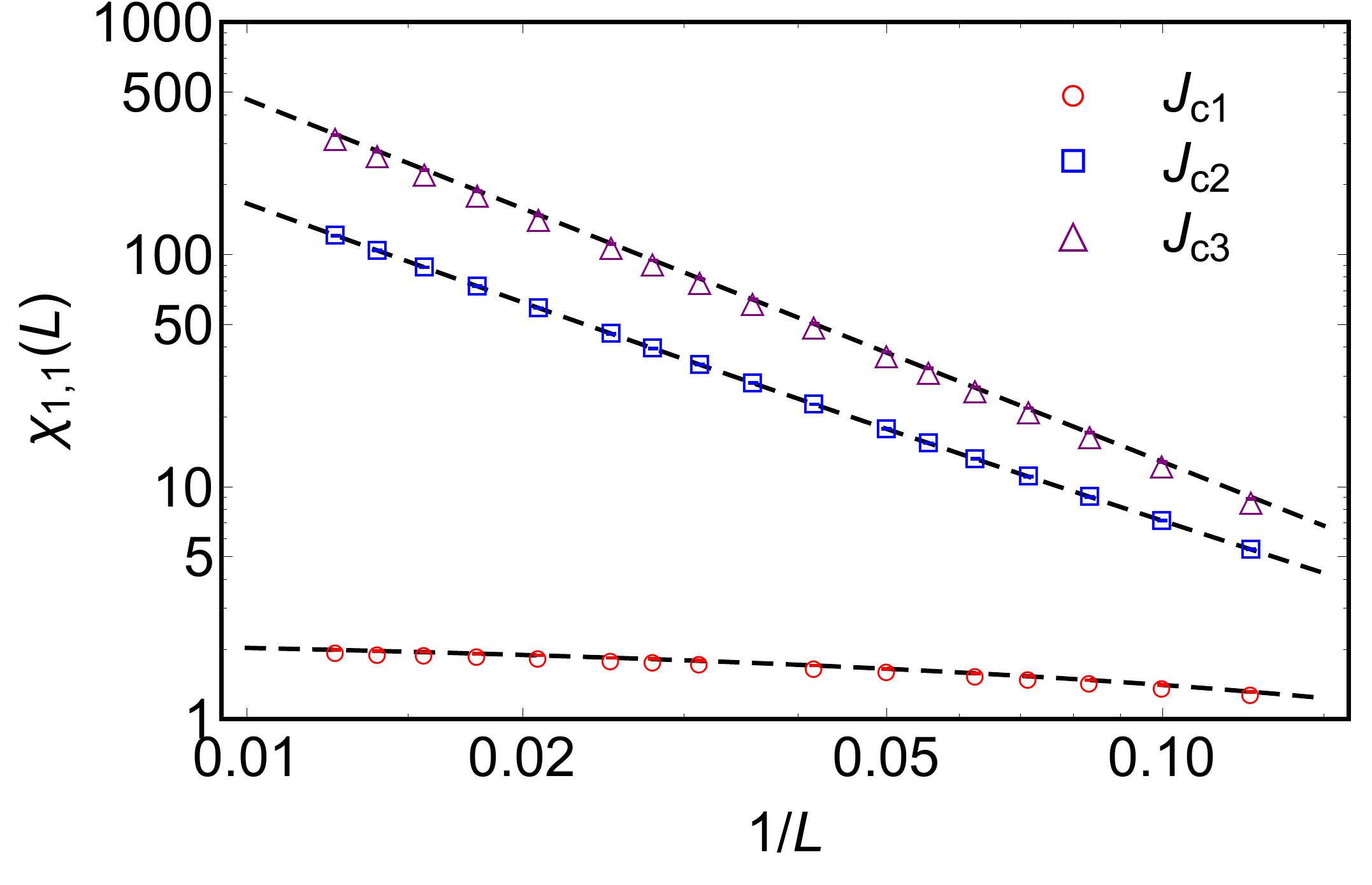}
\caption{The surface staggered susceptibility $\chi_{1,1}$ at the bulk QCPs vs. the inverse lattice size $1/L$. The data at $J_{c1}$ are fitted to $c+aL^{-(2-2y_{h_{1}})}(1+bL^{-1})$, while the data at $J_{c2,3}$ to $aL^{-(2-2y_{h_{1}})}(1+bL^{-1})$. All error bars are much smaller than the symbol sizes.}
\label{fig:chi11}
\end{figure}

\begin{figure}
\centering
\includegraphics[width=0.48\textwidth]{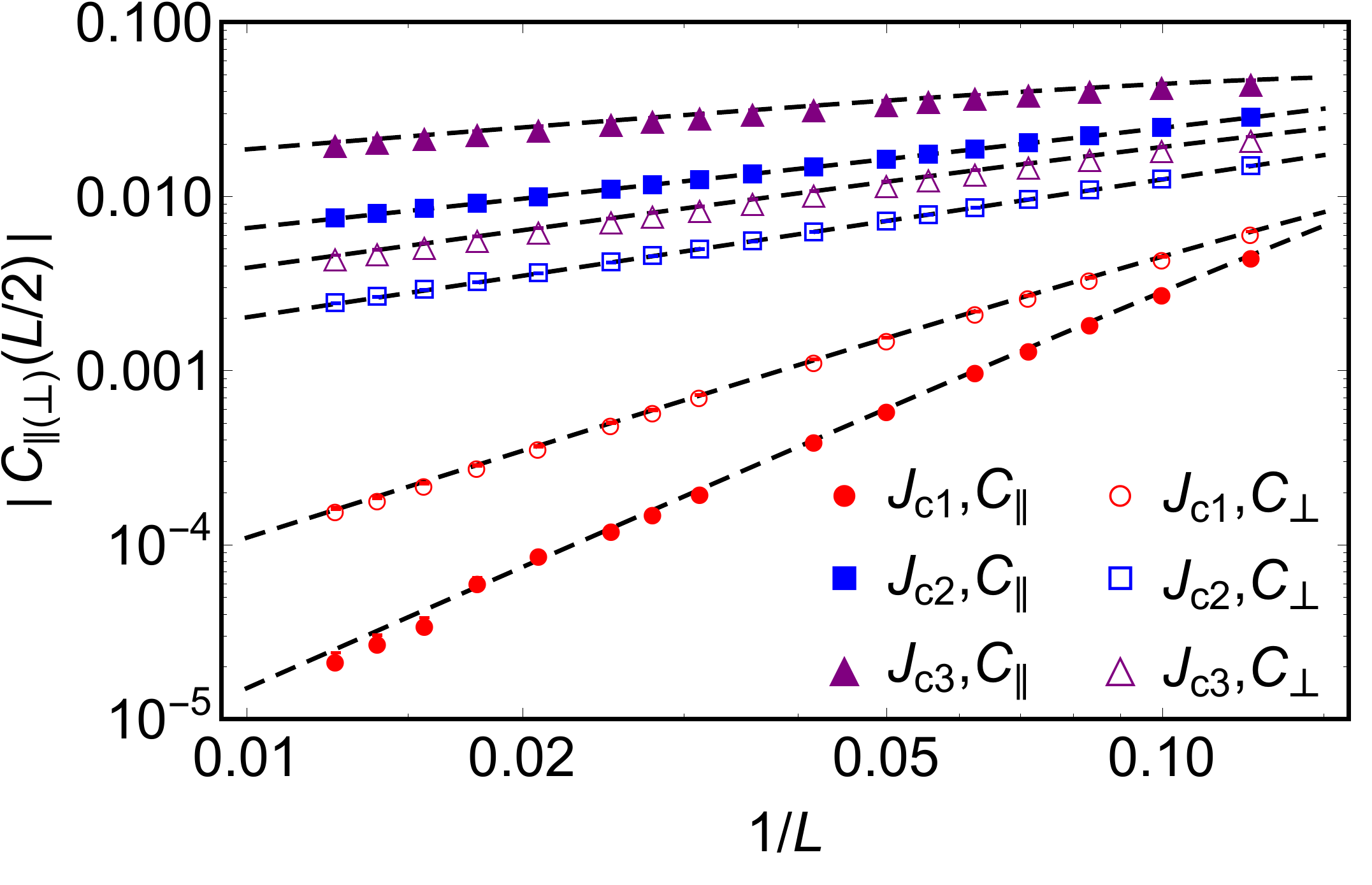}
\caption{The spin correlation functions $C_{\parallel}(L/2)$ and $C_{\perp}(L/2)$ at the bulk QCPs vs. the inverse lattice size $1/L$. These data are fitted to $aL^{-(1+\eta_{\parallel(\perp)})}(1+bL^{-1})$.}
\label{fig:cs}
\end{figure}

On a lattice with an open boundary condition along one direction, we treat the sites that would be dangling bonds in the AKLT state as the surface layers. The surface staggered susceptibility $\chi_{1,1}$ and the spin correlation functions $C_{\parallel}(L/2)$ and $C_{\perp}(L/2)$ are evaluated at the bulk QCPs. The results are shown in Fig. \ref{fig:chi11} and \ref{fig:cs}.

It is easy to see that the surface critical behaviors at $J_{c1}$ are qualitatively different from those at $J_{c2}$ and $J_{c3}$. We first focus on $J_{c1}$. Fitting the $\chi_{1,1}$ data at $J_{c1}$ to Eq. (\ref{eq:chi11}) plus a constant, which takes care of the nonsingular contribution, and a subleading correction term, i.e., to the formula $c+a L^{-(2-2y_{h_{1}})}(1+bL^{-1})$, yields the estimate $y_{h_{1}}=0.810(20)$. It is consistent with the result of the 3D classical Heisenberg model at the ordinary transition \cite{Deng2005}, $y_{h_{1}}=0.813(2)$, thereby confirming the universality of the surface critical behaviors. The spin correlation functions $C_{\parallel}(L/2)$ and $C_{\perp}(L/2)$ are fitted to Eqs. (\ref{eq:cparallel}) and (\ref{eq:cperp}), which yields $\eta_{\parallel}=1.327(25)$, and $\eta_{\perp}=0.680(8)$. These exponents satisfy the scaling relations Eqs. (\ref{eq:relation1}) and (\ref{eq:relation2}) within error bars.

The surface susceptibilities and the correlation functions at $J_{c2}$ and $J_{c3}$ are also fitted to Eqs. (\ref{eq:chi11}), (\ref{eq:cparallel}) and (\ref{eq:cperp}) (see Figs. \ref{fig:chi11} and \ref{fig:cs}). The critical exponents are distinct from those at $J_{c1}$ and in the 3D classical Heisenberg model (Table \ref{tab:exponents}). The scaling dimension $y_{h_{1}}$ is much larger: $y_{h_{1}}=1.7276(14)$ at $J_{c2}$, and $y_{h_{1}}=1.7802(16)$ at $J_{c3}$. The anomalous dimensions are negative: $\eta_{\parallel}=-0.449(5)$ and $\eta_{\perp}=-0.2090(15)$ at $J_{c2}$, and $\eta_{\parallel}=-0.561(4)$ and $\eta_{\perp}=-0.2707(24)$ at $J_{c3}$, suggesting stronger spin correlations on the surface. These exponents also satisfy the scaling relations within error bars; therefore, they consistently establish new universality classes of the surface critical behaviors. Moreover, the exponents at $J_{c2}$ and $J_{c3}$ are slightly different from each other, showing that they belong to two distinct universality classes, which may be a result of the different surface and bulk couplings.

\begin{figure}[t]
\centering
\includegraphics[width=0.48\textwidth]{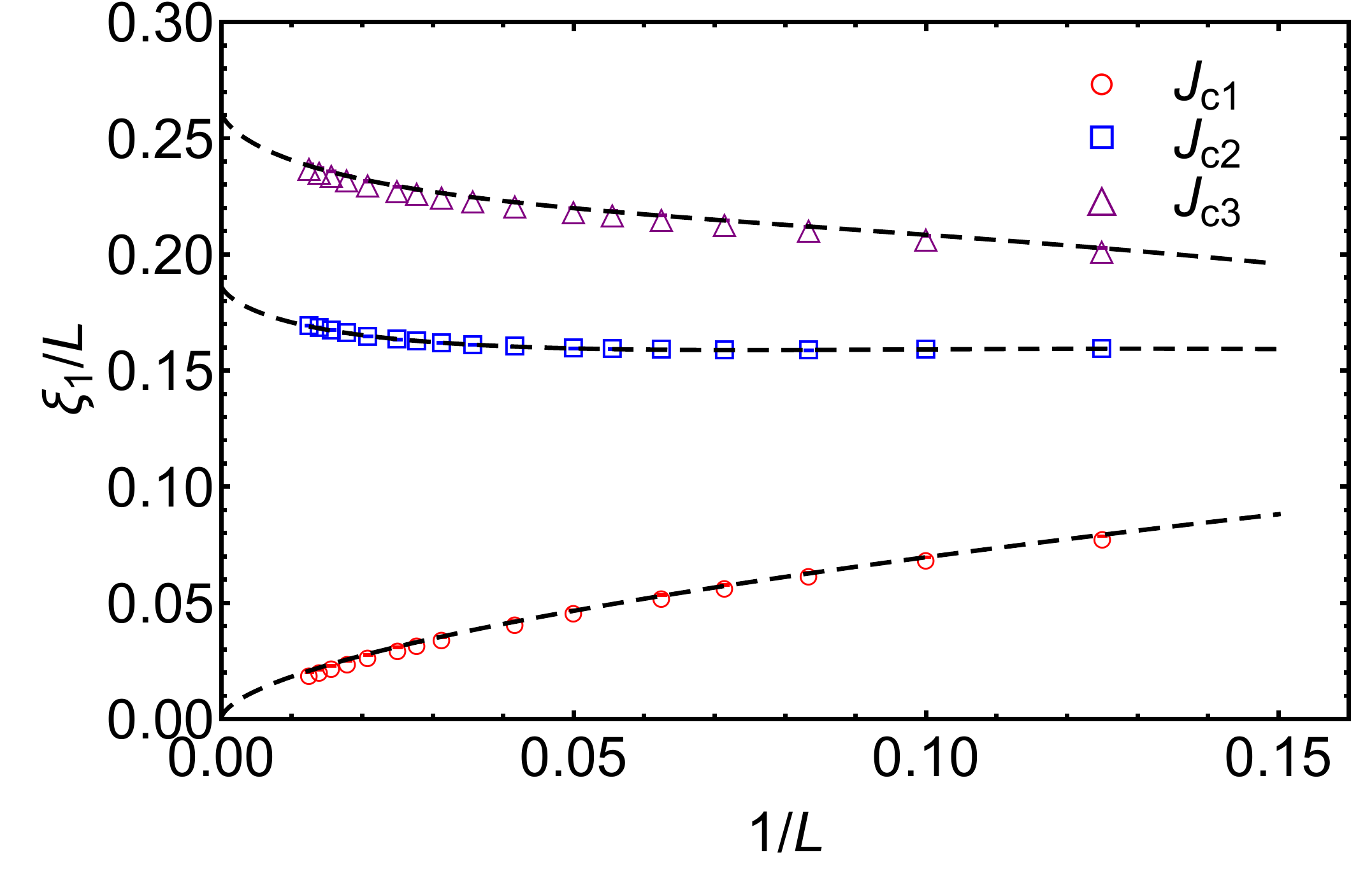}
\caption{The surface spin correlation lengths $\xi_{1}$ divided by the lattice size $L$ vs $1/L$. The data at $J_{c1}$ are fitted to $aL^{-b}$, yielding $b=0.581(4)$, while the data at $J_{c2}$ and $J_{c3}$ are fitted to a constant plus subleading correction terms.}
\label{fig:xis}
\end{figure}

\emph{Discussions.---}In order to highlight the different spin correlations on the surfaces, we calculate the second-moment correlation length on the surface, $\xi_{1}$, at each QCP, which is derived from the surface static structure factors similarly to Eq. (\ref{eq:xi}). As shown in Fig. \ref{fig:xis}, $\xi_{1}/L$ at $J_{c1}$ decays to zero as $L\rightarrow\infty$, which can be fitted with a power law, $\xi_{1}/L\sim L^{-0.581(4)}$. In contrast, $\xi_{1}/L$ at $J_{c2}$ and $J_{c3}$ extrapolate to finite values as $L\rightarrow \infty$. Such a distinction can be phenomenologically understood as follows. At $J_{c1}$, the long-range correlation on the surface is purely induced from the bulk and is much weaker than the bulk correlation. Because the bulk correlation length $\xi$ diverges faster than $\xi_{1}$, the surface critical behaviors are controlled by $\xi$ and the bulk criticality. On the other hand, the gapless surface state in the AKLT phase hybridizes with the bulk critical modes, hence strongly enhancing the surface correlation and resulting in different universality classes of the surface critical behaviors.

The bulk QCPs are described by the 3D O(3) nonlinear $\sigma$ model,
\begin{equation}
L_{b}=\frac{1}{2g}\sum_{\mu=x,y,\tau}(\partial_{\mu}\mathbf{n})^{2},
\end{equation}
which also determines the surface critical behaviors in the conventional case. At QCPs adjacent to the AKLT phase, the surface critical theory should be complemented by the topological $\theta$ term, $\frac{i\theta}{4\pi^{2}}\mathbf{n}\cdot (\partial_{\tau}\mathbf{n}\times \partial_{x}\mathbf{n})$, in which $\mathbf{n}$ is restricted on the surface, and $\theta =\pi$. It originates from the effective spin-$1/2$ Heisenberg chain on the surface and captures the gapless surface state \cite{Haldane1985, Haldane1988}. The detailed theoretical analysis of these unconventional surface critical behaviors is left for future studies.

This physical picture can be generalized to the surface critical behaviors of other SPT phases with gapless surface states at QCPs that spontaneously break the protecting symmetries. The gapless surface states will hybridize with the critical modes and lead to unconventional universality classes of surface critical behaviors. On the other hand, the surface of an SPT phase can be gapped with topological order. In such a case, the critical modes on the surface is carried by fractionalized particles, which is expected to induce an exotic universality class of surface critical behaviors similarly to the QCP from a bulk topological ordered phase to a symmetry-breaking ordered phase \cite{Xu2012, Isakov2012a}.

\emph{Conclusion.---}To summarize, the 2D AKLT phase is realized in the Heisenberg model on a decorated square lattice. Its quantum phase transitions to N\'eel ordered phases are studied in detail with QMC simulations. Although these bulk QCPs belong to the 3D O(3) universality class, which is captured by the Landau phase transition theory, the surface critical behaviors in the presence of free surfaces are shown to belong to different universality classes than the ordinary transition of the 3D classical Heisenberg model. We propose that such a distinction is a physical consequence of the symmetry-protected gapless surface state of the AKLT phase.

\acknowledgements
L.Z. is grateful to Shang-Qiang Ning for enlightening discussions, which partly motivated this work, to Zi-Yang Meng for helpful discussions on the data analysis, and to Shuai Wang for kind help in using the supercomputer. The numerical simulations were performed on Tianhe-I Supercomputer System in Tianjin. This work was supported by the National Key Basic Research Program of China (Grant No. 2014CB920902) and the National Natural Science Foundation of China (Grant No. 11374018).

\bibliography{/Dropbox/ResearchNotes/BibTex/library,/Dropbox/ResearchNotes/BibTex/books}
\end{document}


\title{Supplemental Materials for ``Unconventional Surface Critical Behaviors Induced by Quantum Phase Transition from Two-Dimensional Affleck-Kennedy-Lieb-Tasaki Phase to N\'eel Order''}

\author{Long Zhang}
\affiliation{International Center for Quantum Materials, School of Physics, Peking University, Beijing, 100871, China}

\author{Fa Wang}
\affiliation{International Center for Quantum Materials, School of Physics, Peking University, Beijing, 100871, China}
\affiliation{Collaborative Innovation Center of Quantum Matter, Beijing, 100871, China}

\begin{abstract}
In these Supplemental Materials, the details of the finite-size scaling analysis of the bulk QCPs are presented.
\end{abstract}

\date{\today}
\maketitle


\section{$J_{c1}$: PSL to $S=1/2$ N\'eel order}

\subsection{Quantum critical point}

\begin{figure*}[p]
\centering
\includegraphics[width=0.9\textwidth]{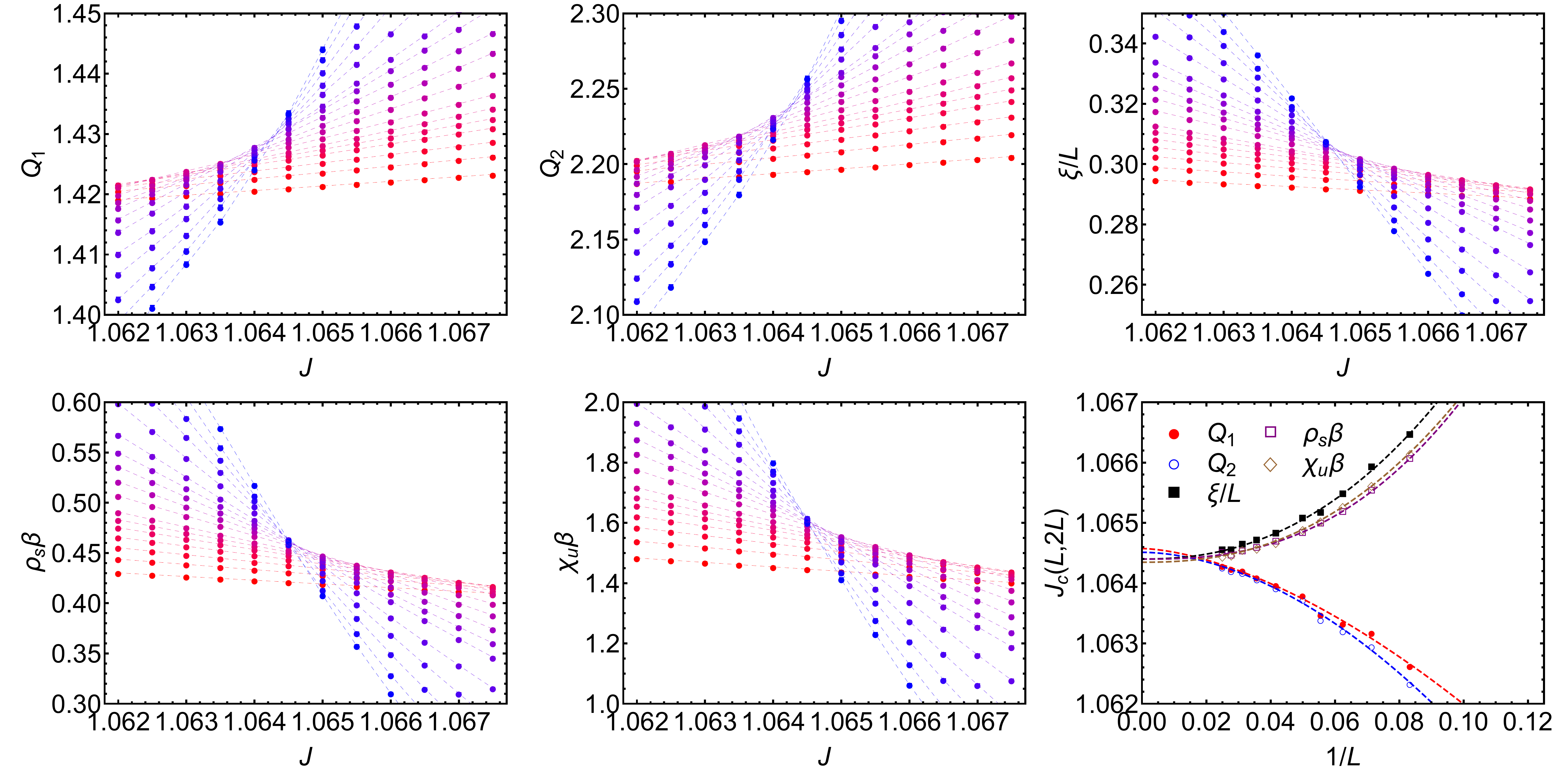}
\caption{Dimensionless quantities measured around $J_{c1}$: Binder ratios $Q_{1}$ and $Q_{2}$, $\xi/L$, $\rho_{s}\beta$, and $\chi_{u}\beta$. Lower right: For each dimensionless quantity, the crossing points $J_{c1}(L,2L)$ are fitted to a power law function, $J_{c1}(L,2L) = J_{c1}+aL^{-\lambda}$. All five quantities yield consistent estimates of $J_{c1}$.}
\label{fig:Jc1}
\end{figure*}

As discussed in the main text, all three QCPs are associated with the spontaneous breaking of the spin rotational symmetry. The order parameter is the staggered magnetization,
\begin{equation}
m_{s}^{z}=\frac{1}{L^{2}}\sum_{i}(-1)^{i}S_{i}^{z},
\end{equation}
in which $(-1)^{i}=+1$ or $-1$ according to the N\'eel order patterns sketched in Fig. 1 of the main text. The Binder ratios are derived from $m_{s}^{z}$, $Q_{1}=\langle (m_{s}^{z})^{2}\rangle / \langle |m_{s}^{z}|\rangle^{2}$, and $Q_{2}=\langle (m_{s}^{z})^{4}\rangle/\langle (m_{s}^{z})^{2}\rangle^{2}$. The second-moment correlation length is derived from the static spin structure factors, $S(\mathbf{q})=\sum_{\mathbf{r}}e^{-i\mathbf{q}\cdot\mathbf{r}}C(\mathbf{r})$,
\begin{equation}
\label{eq:xi}
\xi=\frac{L}{2\pi}\sqrt{\frac{S(\pi,\pi)}{S(\pi+2\pi/L,\pi)}-1},
\end{equation}
in which $C(\mathbf{r})=\langle S_{\mathbf{r}}^{z}S_{0}^{z}\rangle$ is the spin correlation function. The spin stiffness $\rho_{s}$ and the uniform susceptibility $\chi_{u}$ are measured with the improved estimators \cite{Sandvik2010}. At a QCP, $Q_{1}$, $Q_{2}$, and $\xi/L$ are dimensionless, while $\rho_{s}\beta$ and $\chi_{u}\beta$ are also expected to be dimensionless if the QCP has an asymptotic Lorentz invariance with the dynamical critical exponent $z=1$. As shown in Fig. \ref{fig:Jc1}, these curves measured around $J_{c1}$ on different lattice sizes ($L$ from $8$ to $80$) roughly cross at the same point, which confirms that there is a single continuous quantum phase transition and gives a rough estimate of $J_{c1}$.

For each of these dimensionless quantities, the crossing point of the two curves measured on lattice sizes $L$ and $2L$ respectively is denoted by $J_{c1}(L,2L)$. A more precise estimate of $J_{c1}$ is obtained by fitting $J_{c1}(L,2L)$ to a power law, $J_{c1}(L,2L)=J_{c1}+a L^{-\lambda}$. All five quantities yield consistent estimates of $J_{c1}$ (see Fig. \ref{fig:Jc1} and Table \ref{tab:dimensionless}). The weighted average is taken as the final estimate, $J_{c1}=1.064382(13)$, which significantly improves the estimate $1.065(1)$ obtained in previous works \cite{Troyer1997}.

\begin{table*}[t]
\centering
\caption{The estimates of the QCPs and the correlation length exponent $\nu$ at each QCP from the five dimensionless quantities. The weighted averages are taken as the final estimates.}
\label{tab:dimensionless}
\begin{tabular}{ccccccc}
\hline \hline
					& $Q_{1}$			& $Q_{2}$			& $\xi/L$			& $\rho_{s}\beta$	& $\chi_{u}\beta$	& Average			\\
\hline
$J_{c1}$			& $1.06457(12)$		& $1.06451(8)$		& $1.064400(41)$	& $1.064399(20)$	& $1.064347(21)$	& $1.064382(13)$	\\
$\nu$ at $J_{c1}$	& $0.7077(30)$		& $0.7068(32)$		& $0.7062(23)$		& $0.6993(30)$		& $0.7098(30)$		& $0.7060(13)$		\\
$J_{c2}$			& $0.603496(24)$	& $0.603530(16)$	& $0.603548(26)$	& $0.603506(19)$	& $0.603519(33)$	& $0.603520(10)$	\\
$\nu$ at $J_{c2}$	& $0.7221(29)$		& $0.7239(86)$		& $0.7037(14)$		& $0.7038(14)$		& $0.6908(42)$		& $0.7052(9)$		\\
$J_{c3}$			& $-0.934336(34)$	& $-0.934334(32)$	& $-0.934215(14)$	& $-0.934268(30)$	& $-0.934245(47)$	& $-0.934251(11)$	\\
$\nu$ at $J_{c3}$	& $0.7065(49)$		& $0.7032(38)$		& $0.7094(32)$		& $0.7023(28)$		& $0.7057(30)$		& $0.7052(15)$		\\
\hline \hline
\end{tabular}
\end{table*}

\subsection{Critical exponents}

A physical quantity in the quantum critical regime satisfies the following finite-size scaling ansatz,
\begin{equation}
Q(\delta,h;L)\sim L^{\kappa/\nu}\tilde{Q}(\delta L^{y_{\delta}}, hL^{y_{h}}, \alpha_{i}L^{y_{i}}).
\end{equation}
in which $\tilde{Q}$ is a smooth function of its arguments, and $\kappa$ is a critical exponent specific for the quantity. $\delta = J-J_{c1}$. $h$ is the staggered field that couples to the order parameter. $y_{\delta}$ and $y_{h}$ are their scaling dimensions, respectively, which are related to the correlation length exponent $\nu$ and the staggered magnetization exponent $\beta$ by $\nu = 1/y_{\delta}$ and $\beta/\nu = d+z-y_{h}$. $\alpha_{i}$ and $y_{i}$ are the leading-order irrelevant perturbation term and its scaling dimension, $y_{i}<0$. Without a priori knowledge of $y_{i}$, we simply expand the correction factor in terms of a polynomial of $L^{-1}$,
\begin{equation}
\label{eq:fss}
Q(L)|_{\delta=0,h=0} \sim L^{\kappa/\nu}(1+b_{1}L^{-1}+b_{2}L^{-2}+\ldots),
\end{equation}
which works well if $|y_{i}|$ is not too small. In practice, we find that one or two correction terms suffice to yield good fittings to all quantities studied in this work.

\begin{figure*}[p]
\centering
\includegraphics[width=0.6\textwidth]{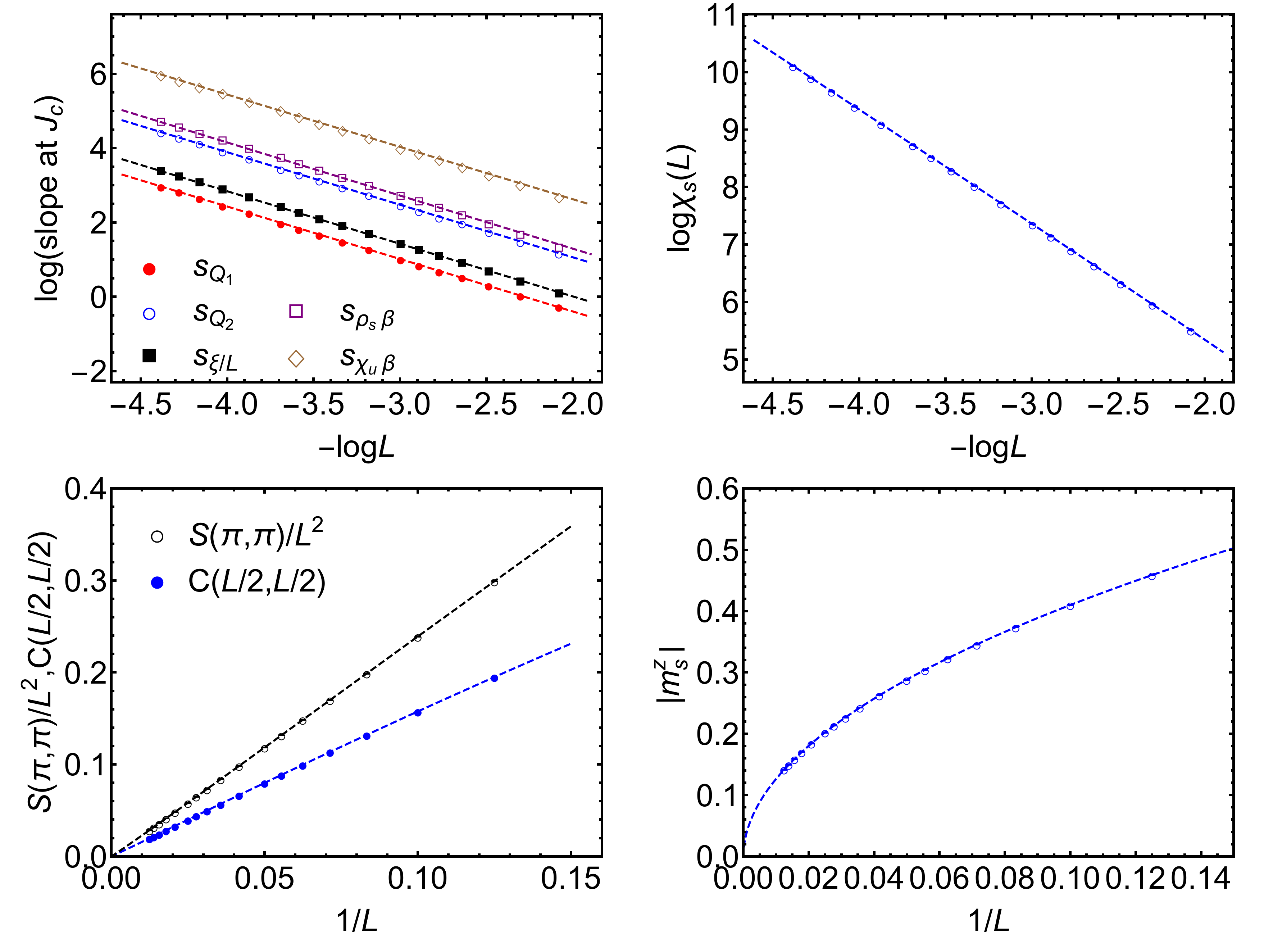}
\caption{Upper left: Slopes of the dimensionless quantities at $J_{c1}$ vs $1/L$ fitted to Eq. (\ref{eq:nu}) in the log-log scale. Upper right: The staggered magnetic susceptibility $\chi_{s}$ at $J_{c1}$ vs $1/L$ in the log-log scale. One correction term in the form of Eq. (\ref{eq:fss}) is included in fitting the anomalous dimension $\eta$. Lower left: The static spin structure factor $S(\pi, \pi)$ divided by $L^{2}$ and the spin correlation function $C(L/2,L/2)$ vs $1/L$. One correction term is included in the fitting. Lower right: The staggered magnetization $|m_{s}^{z}|$ vs $1/L$. One correction term is included in the fitting. The error bars are much smaller than the symbol sizes.}
\label{fig:Jc1exponents}
\end{figure*}

In order to extract the critical exponents, we extrapolate the physical quantities to $J_{c1}=1.0644$ with the reweighting technique. We have confirmed that a slight variation of $J_{c1}$ within its error bar does not affect the estimates of the critical exponents. The scaling ansatz for a dimensionless quantity $Q$ in the quantum critical regime is
\begin{equation}
Q(\delta;L)\sim \tilde{Q}(\delta L^{1/\nu}),
\end{equation}
so the slope of $Q(\delta; L)$ at the QCP, $s_{Q}(L)$, satisfies
\begin{equation}
\label{eq:nu}
s_{Q}(L)\sim L^{1/\nu},
\end{equation}
which is adopted to estimate the exponent $\nu$. The five dimensionless quantities yield consistent results, which is shown in Fig. \ref{fig:Jc1exponents} and Table \ref{tab:dimensionless}. The final estimate is $\nu=0.7060(13)$.

The staggered susceptibility $\chi_{s}$ at the QCP satisfies
\begin{equation}
\chi_{s}(L)\sim L^{2-\eta},
\label{eq:chis}
\end{equation}
in which $\eta$ is the anomalous dimension of the order parameter $m_{s}^{z}$. The fitting shown in Fig. \ref{fig:Jc1exponents} yields $\eta=0.0357(13)$.

The static spin structure factor $S(\pi,\pi)$ and the spin correlation function at half of the lattice size $C(L/2,L/2)$ satisfy
\begin{align}
&S(\pi,\pi)/L^{2} \sim L^{-(z+\eta)},\\
&C(L/2,L/2)\sim L^{-(z+\eta)},
\end{align}
in which $z$ is the dynamical critical exponent. Fitting $S(\pi,\pi)/L^{2}$ yields $z+\eta = 1.0367(10)$, while $C(L/2,L/2)$ yields $z+\eta = 1.0357(17)$, which are consistent with each other. The final estimate is $z+\eta = 1.0365(9)$. Combining it with $\eta$ gives an estimate of $z$, $z = 1.0008(16)$, which is consistent with the asymptotic Lorentz invariance at $J_{c1}$.

The absolute value of the staggered magnetization $|m_{s}^{z}|$ at the QCP has the following finite-size scaling form,
\begin{equation}
|m_{s}^{z}|\sim L^{-\beta/\nu}.
\end{equation}
Fitting the data shown in Fig. \ref{fig:Jc1exponents} gives the estimate $\beta/\nu=0.5188(6)$. Combining it with the estimate of $\nu$ gives $\beta=0.3663(8)$.

The estimates of these critical exponents significantly improve those obtained previously \cite{Troyer1997}, and are consistent with the 3D O(3) universality class \cite{Guida1998}, which is expected from the Landau phase transition theory.

\section{$J_{c2}$ and $J_{c3}$}

\begin{figure*}[p]
\centering
\includegraphics[width=0.9\textwidth]{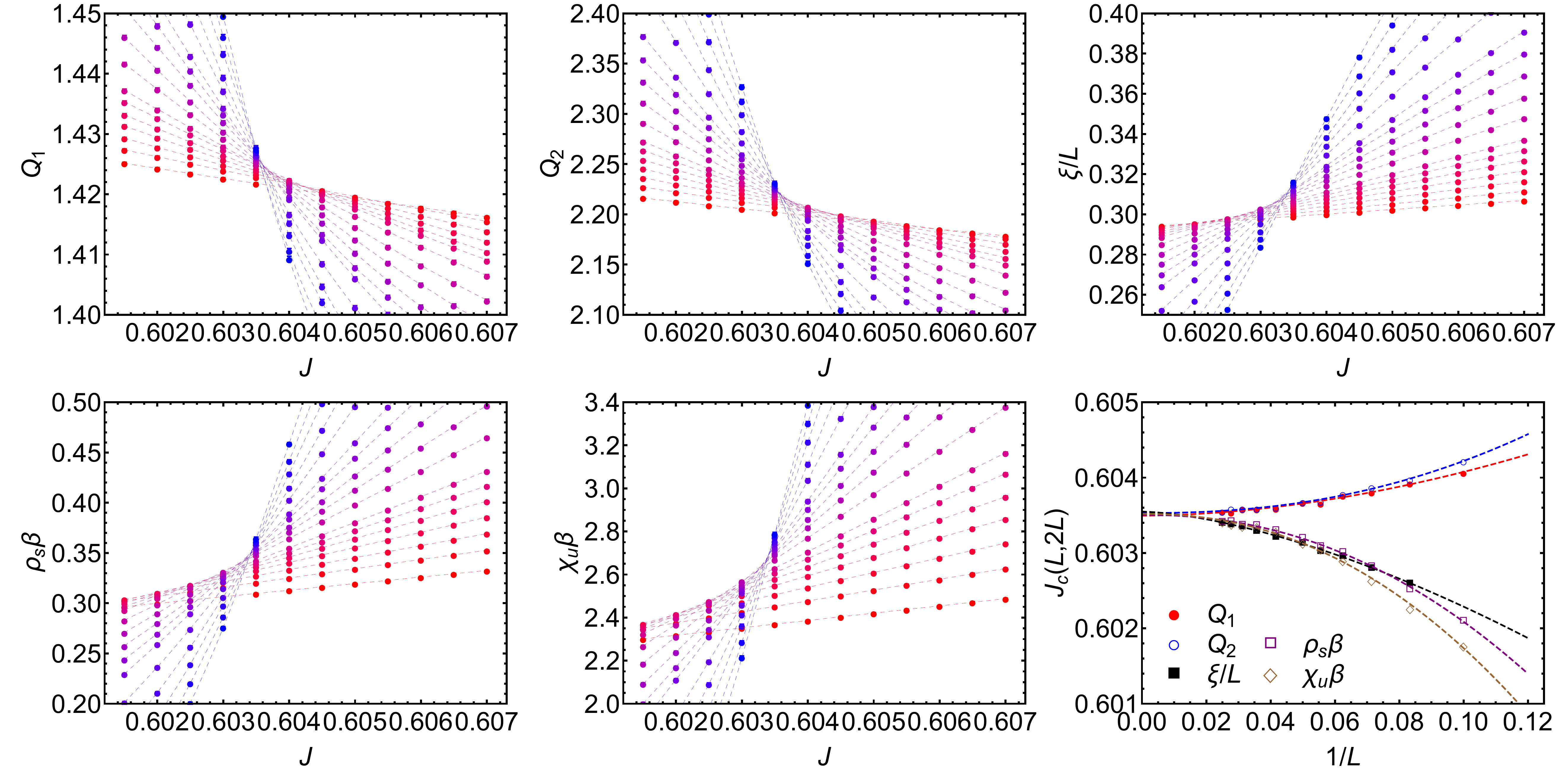}
\caption{Dimensionless quantities measured around $J_{c2}$: Binder ratios $Q_{1}$ and $Q_{2}$, $\xi/L$, $\rho_{s}\beta$, and $\chi_{u}\beta$. Lower right: For each dimensionless quantity, the crossing points $J_{c2}(L,2L)$ are fitted to a power law function, $J_{c2}(L,2L) = J_{c2}+aL^{-\lambda}$. All five quantities yield consistent estimates of $J_{c2}$.}
\label{fig:Jc2}
\end{figure*}

\begin{figure*}[p]
\centering
\includegraphics[width=0.6\textwidth]{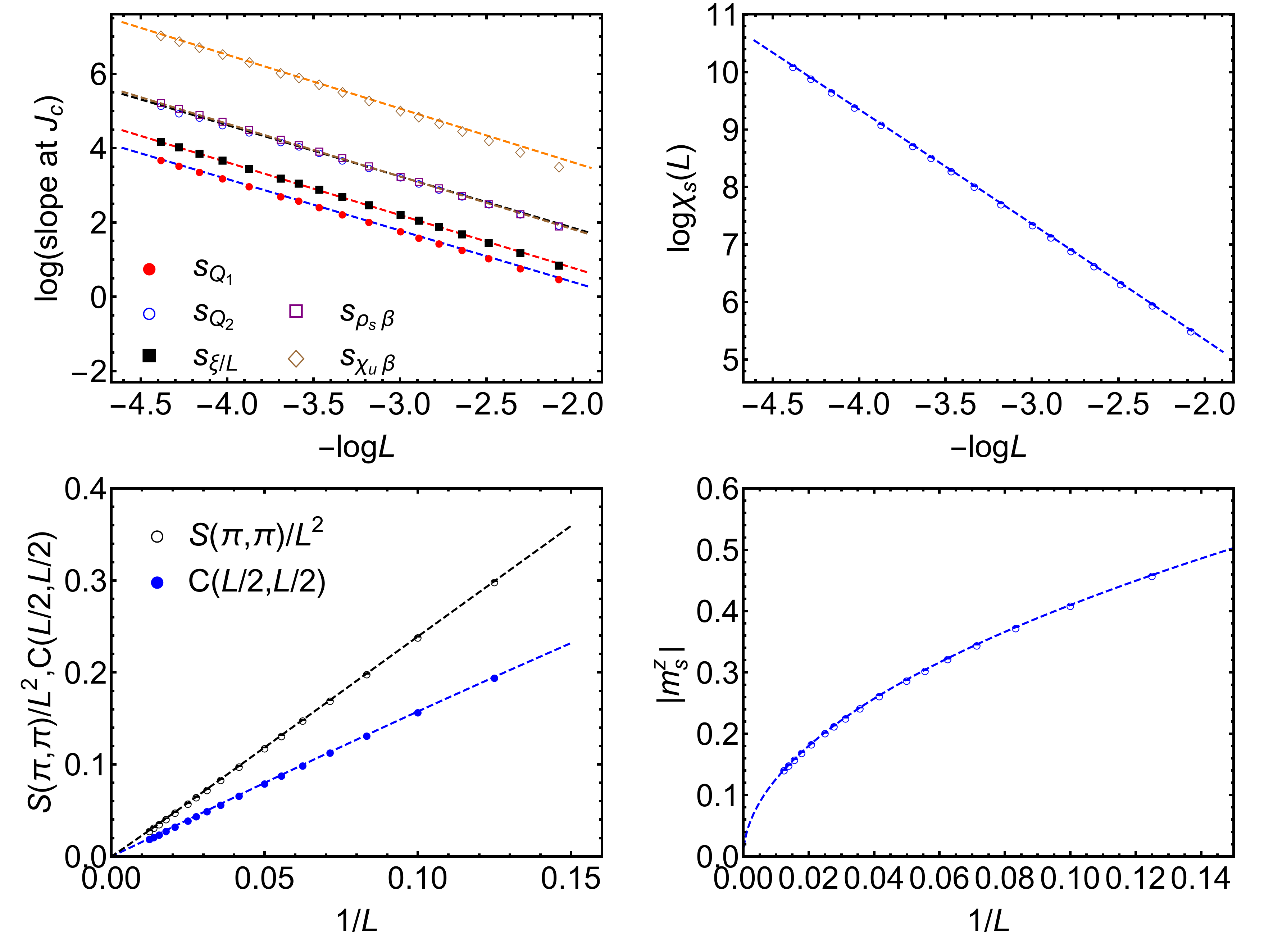}
\caption{Upper left: Slopes of the dimensionless quantities at $J_{c2}$ vs $1/L$ fitted to Eq. (\ref{eq:nu}) in the log-log scale. Upper right: The staggered magnetic susceptibility $\chi_{s}$ at $J_{c2}$ vs $1/L$ in the log-log scale. Two correction terms in the form of Eq. (\ref{eq:fss}) are included in fitting the anomalous dimension $\eta$. Lower left: The static spin structure factor $S(\pi, \pi)$ divided by $L^{2}$ and the spin correlation function $C(L/2,L/2)$ vs $1/L$. Two correction terms are included in the fitting. Lower right: The staggered magnetization $|m_{s}^{z}|$ vs $1/L$. Two correction terms are included in the fitting. The error bars are much smaller than the symbol sizes.}
\label{fig:Jc2exponents}
\end{figure*}

\begin{figure*}[p]
\centering
\includegraphics[width=0.9\textwidth]{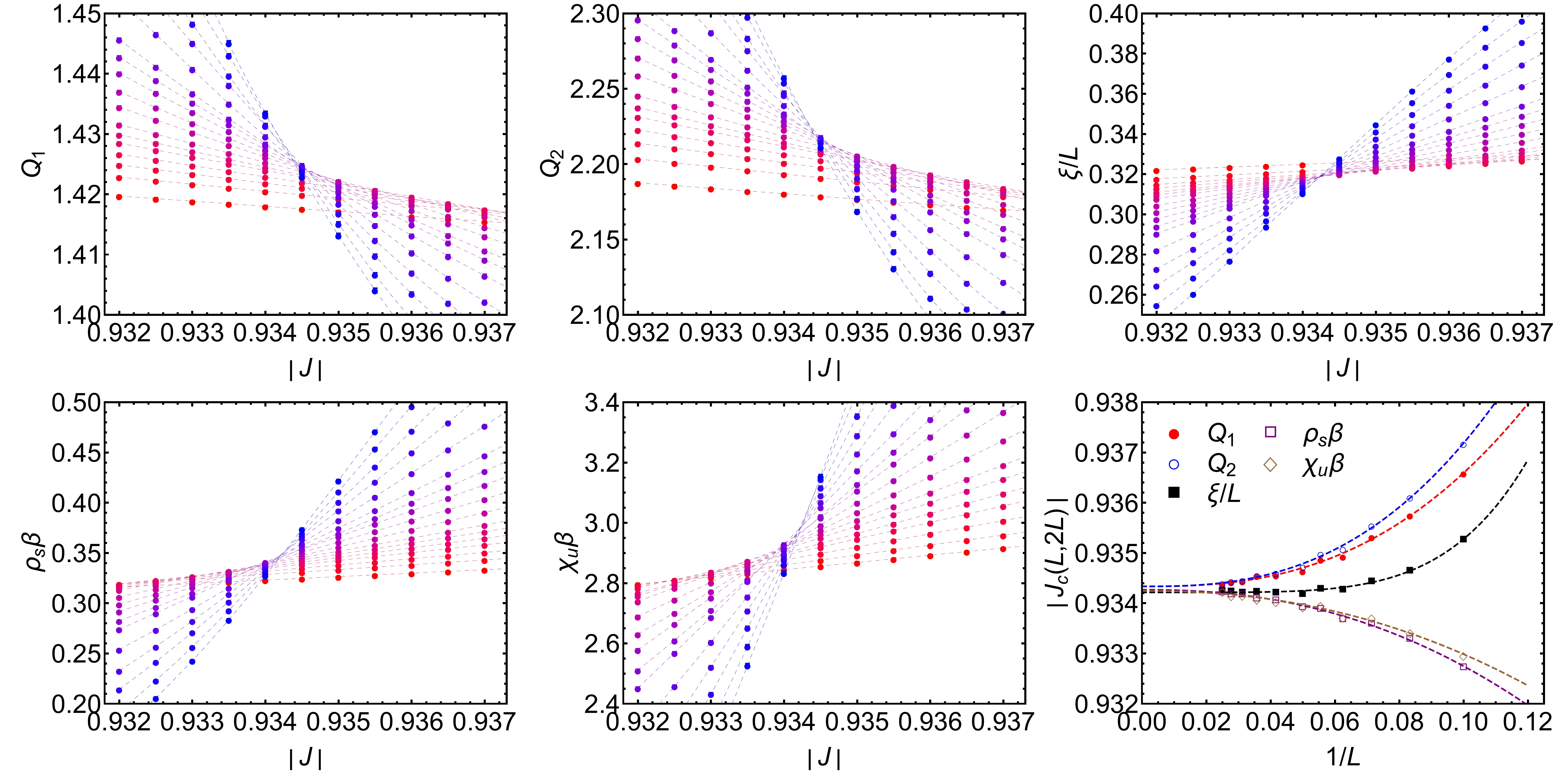}
\caption{Dimensionless quantities measured around $J_{c3}$: Binder ratios $Q_{1}$ and $Q_{2}$, $\xi/L$, $\rho_{s}\beta$, and $\chi_{u}\beta$. Lower right: For each dimensionless quantity, the crossing points $J_{c3}(L,2L)$ are fitted to a power law function, $J_{c3}(L,2L) = J_{c1}+aL^{-\lambda}$. All five quantities yield consistent estimates of $J_{c3}$.}
\label{fig:Jc3}
\end{figure*}

\begin{figure*}[p]
\centering
\includegraphics[width=0.6\textwidth]{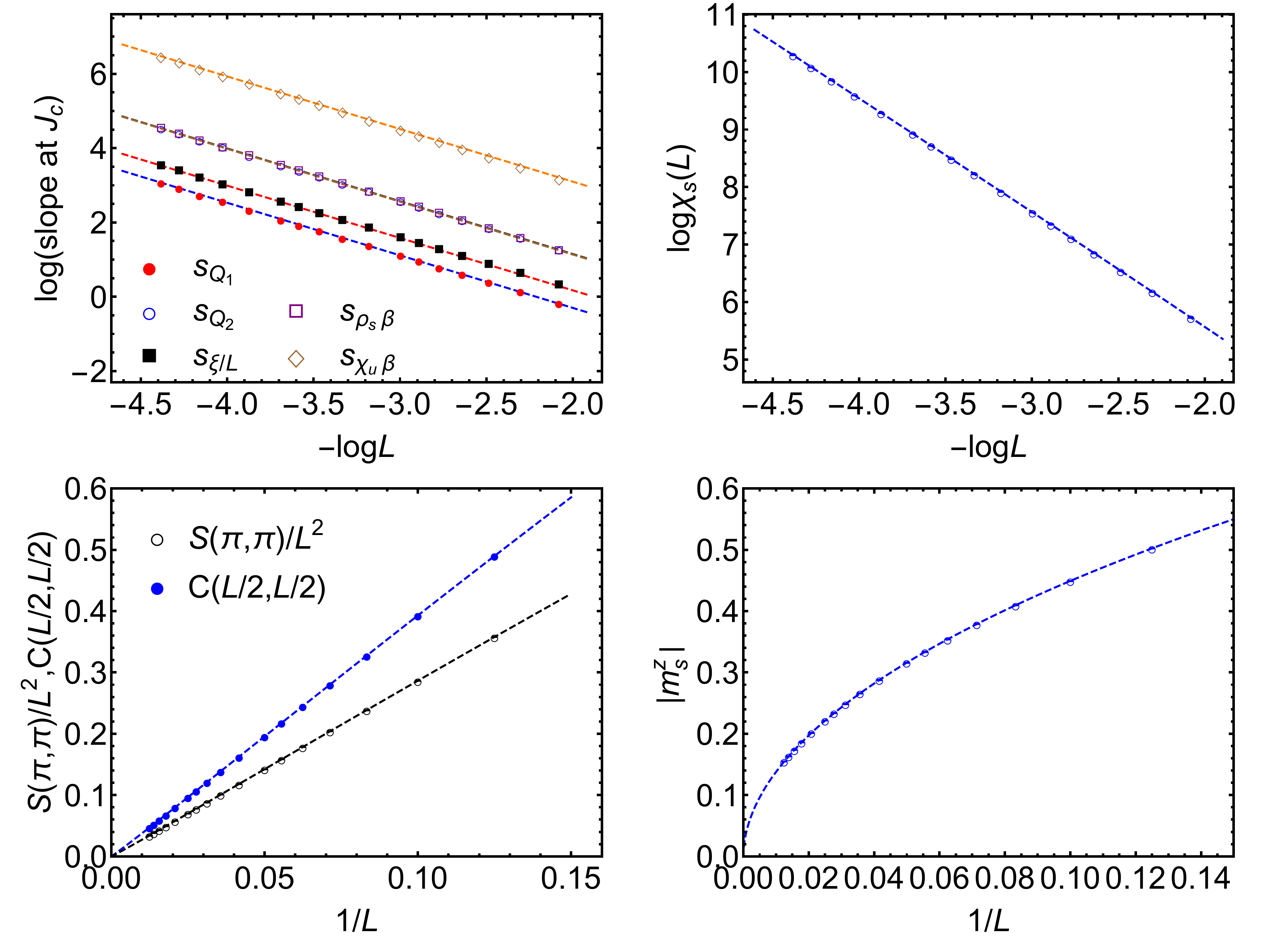}
\caption{Upper left: Slopes of the dimensionless quantities at $J_{c3}$ vs $1/L$ fitted to Eq. (\ref{eq:nu}) in the log-log scale. Upper right: The staggered magnetic susceptibility $\chi_{s}$ at $J_{c3}$ vs $1/L$ in the log-log scale. One correction term in the form of Eq. (\ref{eq:fss}) is included in fitting the anomalous dimension $\eta$. Lower left: The static spin structure factor $S(\pi, \pi)$ divided by $L^{2}$ and the spin correlation function $C(L/2,L/2)$ vs $1/L$. One correction term is included in the fitting. Lower right: The staggered magnetization $|m_{s}^{z}|$ vs $1/L$. One correction term is included in the fitting. The error bars are much smaller than the symbol sizes.}
\label{fig:Jc3exponents}
\end{figure*}

The same finite-size scaling analysis is applied to the other two QCPs, which are from the AKLT phase to the N\'eel ordered phases. The results are summarized in Figs. \ref{fig:Jc2}--\ref{fig:Jc3exponents}, and in Table \ref{tab:dimensionless} and Table I of the main text.

\bibliography{/Dropbox/ResearchNotes/BibTex/library,/Dropbox/ResearchNotes/BibTex/books}